\documentclass[aps,prl,preprint,tightenlines,superscriptaddress,showpacs,byrevtex,subfigure]{revtex4}

\usepackage{graphicx}
\usepackage{color}

\begin{document}
\begin{flushright}
KEK Preprint 2011-31\\
Belle Preprint 2012-8\\
\end{flushright}
\title{ \quad\\[0.5cm] 
Search for Lepton-Flavor and 
Lepton-Number-Violating
$\tau \to \ell hh'$ Decay Modes
}

\begin{abstract}

We search for lepton-flavor and lepton-number-violating
$\tau$ decays 
into a lepton ($\ell =$ electron or muon) and two charged mesons
($h, h' = \pi^{\pm}$ or $K^\pm$)
using
854 fb$^{-1}$ of data collected 
with the Belle detector at the 
KEKB asymmetric-energy $e^+e^-$ collider. 
We obtain 
90\% {confidence level}  upper limits on the 
{$\tau\to \ell hh'$}
branching fractions in the {range}
$(2.0-8.4)\times 10^{-8}$.
These results improve 
{upon our previously} 
published upper limits by factors {of} {about} 1.8 on  average.

\end{abstract}
\affiliation{University of Bonn, Bonn, Germany}
\affiliation{Budker Institute of Nuclear Physics SB RAS and Novosibirsk State University, Novosibirsk 630090, Russian Federation}
\affiliation{Faculty of Mathematics and Physics, Charles University, Prague, The Czech Republic}
\affiliation{Department of Physics, Fu Jen Catholic University, Taipei, Taiwan}
\affiliation{Hanyang University, Seoul, South Korea}
\affiliation{University of Hawaii, Honolulu, HI, USA}
\affiliation{High Energy Accelerator Research Organization (KEK), Tsukuba, Japan}
\affiliation{Indian Institute of Technology Guwahati, Guwahati}
\affiliation{Institute of High Energy Physics, Chinese Academy of Sciences, Beijing, PR China}
\affiliation{Institute for High Energy Physics, Protvino, Russian Federation}
\affiliation{Institute of High Energy Physics, Vienna, Austria}
\affiliation{Institute for Theoretical and Experimental Physics, Moscow, Russian Federation}
\affiliation{J. Stefan Institute, Ljubljana, Slovenia}
\affiliation{Kanagawa University, Yokohama, Japan}
\affiliation{Institut f\"ur Experimentelle Kernphysik, Karlsruher Institut f\"ur Technologie, Karlsruhe, Germany}
\affiliation{Korea Institute of Science and Technology Information, Daejeon, South Korea}
\affiliation{Korea University, Seoul, South Korea}
\affiliation{Kyungpook National University, Taegu, South Korea}
\affiliation{\'Ecole Polytechnique F\'ed\'erale de Lausanne, EPFL, Lausanne, Switzerland}
\affiliation{Faculty of Mathematics and Physics, University of Ljubljana, Ljubljana, Slovenia}
\affiliation{University of Maribor, Maribor, Slovenia}
\affiliation{Max-Planck-Institut f\"ur Physik, M\"unchen, Germany}
\affiliation{University of Melbourne, Victoria, Australia}
\affiliation{Graduate School of Science, Nagoya University, Nagoya, Japan}
\affiliation{Kobayashi-Maskawa Institute, Nagoya University, Nagoya, Japan}
\affiliation{Nara Women's University, Nara, Japan}
\affiliation{National Central University, Chung-li, Taiwan}
\affiliation{National United University, Miao Li, Taiwan}
\affiliation{Department of Physics, National Taiwan University, Taipei, Taiwan}
\affiliation{H. Niewodniczanski Institute of Nuclear Physics, Krakow, Poland}
\affiliation{Nippon Dental University, Niigata, Japan}
\affiliation{Niigata University, Niigata, Japan}
\affiliation{University of Nova Gorica, Nova Gorica, Slovenia}
\affiliation{Osaka City University, Osaka, Japan}
\affiliation{Pacific Northwest National Laboratory, Richland, WA, USA}
\affiliation{Panjab University, Chandigarh, India}
\affiliation{Research Center for Nuclear Physics, Osaka University, Osaka, Japan}
\affiliation{University of Science and Technology of China, Hefei, PR China}
\affiliation{Seoul National University, Seoul, South Korea}
\affiliation{Sungkyunkwan University, Suwon, South Korea}
\affiliation{School of Physics, University of Sydney, NSW 2006, Australia}
\affiliation{Tata Institute of Fundamental Research, Mumbai, India}
\affiliation{Excellence Cluster Universe, Technische Universit\"at M\"unchen, Garching, Germany}
\affiliation{Toho University, Funabashi, Japan}
\affiliation{Tohoku Gakuin University, Tagajo, Japan}
\affiliation{Tohoku University, Sendai, Japan}
\affiliation{Department of Physics, University of Tokyo, Tokyo, Japan}
\affiliation{Tokyo Institute of Technology, Tokyo, Japan}
\affiliation{Tokyo Metropolitan University, Tokyo, Japan}
\affiliation{Tokyo University of Agriculture and Technology, Tokyo, Japan}
\affiliation{CNP, Virginia Polytechnic Institute and State University, Blacksburg, VA, USA}
\affiliation{Yamagata University, Yamagata, Japan}
\affiliation{Yonsei University, Seoul, South Korea}
\author{Y.~Miyazaki} 
\affiliation{Graduate School of Science, Nagoya University, Nagoya, Japan}
\author{K.~Hayasaka} 
\affiliation{Kobayashi-Maskawa Institute, Nagoya University, Nagoya, Japan}
\author{I.~Adachi} 
\affiliation{High Energy Accelerator Research Organization (KEK), Tsukuba, Japan}
\author{H.~Aihara} 
\affiliation{Department of Physics, University of Tokyo, Tokyo, Japan}
\author{D.~M.~Asner} 
\affiliation{Pacific Northwest National Laboratory, Richland, WA, USA}
\author{V.~Aulchenko} 
\affiliation{Budker Institute of Nuclear Physics SB RAS and Novosibirsk State University, Novosibirsk 630090, Russian Federation}
\author{T.~Aushev} 
\affiliation{Institute for Theoretical and Experimental Physics, Moscow, Russian Federation}
\author{A.~M.~Bakich} 
\affiliation{School of Physics, University of Sydney, NSW 2006, Australia}
\author{A.~Bay} 
\affiliation{\'Ecole Polytechnique F\'ed\'erale de Lausanne, EPFL, Lausanne, Switzerland}
\author{V.~Bhardwaj} 
\affiliation{Nara Women's University, Nara, Japan}
% ...
\collaboration{The Belle Collaboration}
\noaffiliation

\maketitle
 \section{Introduction}

{Lepton flavor violation (LFV)
in charged lepton decays is forbidden 
{in the Standard Model (SM)} {and highly suppressed} 
{even if neutrino mixing is taken into account.
On the other hand,}
extensions of the SM,
such as supersymmetry, leptoquark and many other 
models~\cite{cite:amon,cite:six_fremionic,cite:susy1,cite:susy2,Benbrik:2008ik,Li:2009yr,Li:2010vf,Liu:2009su},
predict LFV with  branching fractions
as {large} as {$10^{-8}$,}
{which 
{are}
accessible
in
current
$B$-factory experiments.}
{We search for neutrinoless 
lepton-flavor-violating $\tau^- \to \ell^- h^+ h'^-$ decays 
and lepton-number-violating $\tau^- \to \ell^+ h^- h'^-$ 
decays$^{\footnotemark[1]}$,
where $\ell$ is an electron or muon and $h^{(')}$ is a charged pion or kaon. 
We analyse a 
{854 fb$^{-1}$ data sample} collected with 
the Belle detector~\cite{Belle} at the KEKB
asymmetric-energy   $e^+e^-$ collider~\cite{kekb}
at center-of-mass (CM) energies {at or below}
the $\Upsilon(4S)$ and $\Upsilon(5S)$ resonances.}
Previously, we
obtained
90\% confidence level (C.L.) upper limits
{on} {the} 
branching fractions
using 671 fb${}^{-1}$ of data;
the results were
in the range (3.3$-$16)~$\times~10^{-8}$~\cite{lhh_belle}.
The BaBar collaboration
has
also
published 
{90\% C.L. upper limits} 
in the range (7$-$48)~$\times~10^{-8}$
using 221 fb${}^{-1}$ of data~\cite{lhh_babar}.


{\footnotetext[1]{Throughout this paper,
charge-conjugate modes are 
{implied;} {hence,} the  
notation $\tau \to \ell h h'$ includes 
both $\tau^- \to \ell^- h^+ h'^-$ and $\tau^- \to \ell^+ h^- h'^-$ modes.}}


The Belle detector is a large-solid-angle magnetic spectrometer that
consists of a silicon vertex detector (SVD), 
a 50-layer central drift chamber (CDC), 
an array of aerogel threshold 
{{C}herenkov} counters (ACC), a barrel-like arrangement of 
time-of-flight scintillation counters (TOF), and an electromagnetic calorimeter 
comprised of  
CsI(Tl) {crystals (ECL), all located} inside
a superconducting solenoid coil
that provides a 1.5~T magnetic field.  
An iron flux-return located outside of the coil is instrumented to detect 
$K_{\rm{L}}^0$ mesons 
and to identify muons (KLM).  
The detector is described in detail elsewhere~\cite{Belle}.

Particle identification
is very important
{for this measurement.}
{We use 
{particle} identification likelihood variables} 
based on
the ratio of the energy
deposited in the
ECL to the momentum measured in the SVD and CDC,
shower shape in the ECL,
the {particle's} range in the KLM,
hit information from the ACC,
{$dE/dx$ {measured} in the CDC,}
and {the {particle's time of flight}}.
To distinguish hadron species,
we use likelihood ratios,
${\cal{P}}(i/j) = {\cal{L}}_i/({\cal{L}}_i + {\cal{L}}_{j})$,
where ${\cal{L}}_{i}$ (${\cal{L}}_{j}$)
is the likelihood {of} the {observed} 
detector response
{for} {a} track with flavor $i$ ($j$).
{{For lepton identification,
we {form} likelihood ratios ${\cal P}(e)$~\cite{EID}
and ${\cal P}({\mu})$~\cite{MUID}}
{using}
the responses of the appropriate subdetectors.

{We use {Monte Carlo} (MC) samples}
to estimate the signal efficiency and 
optimize the event selection.
{Signal} and background {event samples} 
from generic $\tau^+\tau^-$ decays are 
generated by KKMC/TAUOLA~\cite{KKMC}. 
{For signal,} 
we generate {the}
{$e^+e^-\to\tau^+\tau^-$} {process,} 
{in which} 
{one} $\tau$ {is forced to decay}  
into  {a lepton} 
and two charged {mesons} 
{using {a three-body {phase space} model,}}
{while} 
the other $\tau$ decays 
{following SM branching ratios.}
{Background {samples from}
$B\bar{B}$ and continuum $e^+e^-\to q\bar{q}$ ($q=u,d,s,c$) 
processes are generated by EvtGen~\cite{evtgen}
while 
Bhabha and two-photon processes are generated by 
BHLUMI~\cite{BHLUMI}
and
{AAFH~\cite{AAFH}}, respectively. }
{In what follows,}
all kinematic variables are calculated in the laboratory frame
unless otherwise specified.
In particular,
variables
calculated in the $e^+e^-$ {CM frame}
are indicated by the superscript {``CM.''}

\section{Event Selection}

%
%

{We search for $\tau^+\tau^-$ events in which one $\tau$ 
(the {signal $\tau$}) decays into a lepton
and
two charged  mesons  ($h, h' =\pi^\pm$ or $K^\pm$), 
and the other $\tau$~(the {tag $\tau$}) 
decays 
into  one charged track, any number of additional 
photons and neutrinos.

{For each candidate event we calculate}
the $\ell hh'$ invariant
mass~($M_{ {\ell hh'}}$) {and} 
the difference of {the $\ell hh'$} energy from the 
beam energy in the CM {frame}~($\Delta E$).
{In the 
two-dimensional distribution of $M_{{\ell hh'}}$ versus $\Delta E$,}
{signal events} 
should have $M_{{\ell hh'}}$
close to the $\tau$-lepton mass ($m_{\tau}$) and
$\Delta E$ close to zero.
{From MC {simulations,}
{the dominant
background}
for the $\tau\to \mu hh'$ {modes} 
{is} from
continuum and
generic
$\tau^+\tau^-$ {processes,}
while {that}
for the $\tau\to ehh'$ modes
{is}
from two-photon processes.}
Therefore, 
the event selection is optimized mode-by-mode
since the backgrounds are mode dependent.
This analysis includes 27\% more data than 
in the previous one~\cite{lhh_belle}.
We reoptimized the selection
criteria by reblinding the whole
data set in the signal region until all
selection criteria are finalized. The selection criteria are determined
so that the figure of merit (FOM), 
defined by FOM=(Signal detection efficiency)/$\sqrt{\mbox{No. of
expected background events}}$, is maximized.
Here, the expected background
is estimated using events from
the simulated samples
within the $\pm20\sigma_{M_{\ell h h'}}$ 
and $\pm5\sigma_{\Delta E}$ region
on the $M_{\ell h h'}$ vs $\Delta E$ plane,
where $\sigma_{M_{\ell h h'}/\Delta E}$ is
the resolution of $M_{\ell h h'}$ or $\Delta E$, respectively, and 
details will be discussed later.

%
%

Candidate $\tau$-pair events are required to have 
four tracks  with {zero} net charge.
{All charged tracks} and photons 
are required to be reconstructed 
{within {the} fiducial {volume}} 
defined by $-0.866 < \cos\theta < 0.956$,
where $\theta$ is the polar angle with
{respect to the direction 
{along}
the $e^+$ beam.}
{Each charged track should have transverse momentum ($p_t$)
greater than 0.1 GeV/$c$
while each photon should have energy ($E_{\gamma}$)
greater than 0.1 GeV.}
{For each charged track, 
the distance of the closest point with 
respect to the interaction point 
is required to be 
less than 0.5 cm in the transverse direction 
and less than 
3.0 cm in the longitudinal direction.}

%
%

{Using the plane perpendicular to the CM
thrust axis~\cite{thrust},
which is calculated from 
the observed tracks and photon candidates,
we separate the particles in an event
into two hemispheres.
These  are referred to as the signal and 
tag sides. }
{The tag side is required to  
{include a single charged track}
while the signal side is required to 
{contain three charged tracks.}}
{We require one of three 
charged tracks on the signal side 
{to be} 
identified as a lepton.}
{The electron and muon {identification} criteria are}
${\cal P}(e) > 0.9$ with 
{momentum} $p > 0.6$ GeV/$c$
and 
${\cal P}(\mu) > 0.95$ with $p >1.0 $ GeV/$c$,
respectively.
In order to take into account the emission
of  bremsstrahlung photons from the electron,
the momentum of the
electron {candidate}
is reconstructed 
by adding to it the
momentum of every photon
within 0.05 radians of
the electron track direction.
The electron (muon) identification
efficiency is 89\% (81\%) while
the probability to misidentify a pion
as an electron (a muon) is 0.7\% (1.2\%).

%
%

{Charged kaons are} identified by
a condition ${\cal{P}}(K/\pi) > (0.6-0.9$) 
for each mode, 
{as shown {in  Table~\ref{tbl:thrust}},}
while {charged pions are} identified by
{the requirement}
${\cal{P}}(K/\pi) < 0.6$.
Furthermore,
we apply {a} proton veto for kaon 
{candidates, ${\cal{P}}(p/K)<0.6$,}
to reduce 
{protons incorrectly identified as kaons.}
The kaon (pion) identification
{efficiency is} around 80\% (88\%)
while
{the probability to misidentify {a} pion
(kaon)
as
{a} {kaon} (pion)}
is below 10\% (12\%).
{{In order to {suppress}
background
from {photon} conversions
({\it i.e.} $\gamma \rightarrow e^+e^-$),}
we require 
{each $h$ or $h'$ candidate track} to have
${\cal{P}}(e) <0.1$.}
Furthermore, we {apply  the condition}
${\cal{P}(\mu)} <0.1$
to suppress two-photon background 
{from} $e^+e^- \to e^+e^-\mu^+\mu^-$. 

\begin{table}
\caption{Selection criteria {for}
kaon identification ${\cal{P}}(K/\pi)$ 
and magnitude of thrust ($T$).}
\label{tbl:thrust}
\begin{tabular}{|c|c|c|}\hline \hline 
Mode & {${\cal{P}}(K/\pi)$}  & $T$  \\\hline
$\tau\to\mu\pi\pi$ &  $-$ &
$0.90 < T <0.98 $ \\
$\tau\to\mu K\pi$ &  $>0.9$  &
$0.92 < T < 0.98$ \\
$\tau\to\mu KK$ &  $>0.8$  &
$0.92 < T < 0.98$ \\
$\tau\to e\pi\pi$ &  $-$   &
$0.90 < T < 0.97$ \\
$\tau\to eK\pi$  & $>0.8$  &
$0.90 < T < 0.97$ \\
$\tau\to eKK$  & $> 0.6$  &
$0.90 < T < 0.98$ \\\hline\hline
\end{tabular}
\end{table}

%
%

Because a signal decay has no neutrino,
the missing momentum is entirely due 
to neutrinos emitted from the tag side.
The missing
momentum $\vec{p}_{\rm miss}$ 
is defined 
as a difference between
the sum of the $e^+$ and $e^-$ beam momenta and
the vector sum of the momenta of all tracks and photons,
where photons are
measured in the ECL
as clusters  to which no charged tracks are associated.
Since the direction of $\vec{p}_{\rm miss}$
should lie within the tag side of the event,
the cosine of the opening angle between
$\vec{p}_{\rm miss}$ and the charged track on the tag side 
in the CM system, $\cos \theta^{\mbox{\rm \tiny CM}}_{\rm tag-miss}$, 
is required to be greater than zero.
If the track on the tag side is a hadron,
we also require 
$\cos \theta^{\mbox{\rm \tiny CM}}_{\rm tag-miss}<0.85$ 
for the $\tau\to\mu hh'$ modes.
This requirement reduces continuum background 
with missing energy due to neutrons or $K^0_L$'s
since the masses of the neutron and $K^0_L$ are larger
than that of the neutrino.
We also require that $\cos \theta^{\mbox{\rm \tiny CM}}_{\rm tag-miss}<0.96$ 
for $\tau\to e hh'$ modes.
This requirement reduces Bhabha, 
inelastic vector meson-photoproduction,
and two-photon background,
since these processes produce
electron in the tag-side in many cases and
the electron can produce the radiated photons which
result in missing momentum if they overlap with
the ECL clusters associated with
the tag-side track~\cite{cite:tau_egamma}.
In addition, in order to ensure that the missing particles are neutrinos
rather than photons or charged particles that 
leave the detector acceptance,
we impose requirements on $\vec{p}_{\rm miss}$:
We require that $|p^{\rm{t}}_{\rm miss}|$, 
the magnitude of the transverse component of
$\vec{p}_{\rm miss}$,
be  greater than 0.5 GeV/$c$ (0.7 GeV/$c$) for {the} 
$\tau\to\mu hh'$ ($ehh'$) modes,
and that its direction 
point into the fiducial volume of the
detector.
For the $\tau^-\to e^-\pi^+\pi^-$ mode only,
we apply the tighter selection requirement 
$|{p}^{\rm t}_{\rm miss}|$  $>$ 1.5 GeV/$c$.

To reject continuum, Bhabha and $\mu^+\mu^-$ background,
we require the magnitude of the thrust ($T$)
to be in the ranges
{given} in Table \ref{tbl:thrust}.

To suppress the $B\bar{B}$ and continuum background,
{we require}
that 
the number of photons on the tag side 
{be $n_{\gamma}^{\rm{TAG}} \le 2$ 
and $n_{\gamma}^{\rm{TAG}}\le 1$}
{for decays with hadronic and leptonic {tags}, respectively.}
A leptonic tag is defined as ${\cal P}(e) > 0.1$ or ${\cal P}(\mu) > 0.1$ 
while a tag is {hadronic} 
if the {leptonic requirements are not satisfied.}
{We allow at most one additional 
photon on the signal side.}
{The reconstructed mass of {the tag side,} 
{combining the photons with the charged track
(assumed to be a pion mass)}
from the tag side, $m_{\rm tag}$, is
required to be less than {1.0 GeV/$c^2$ 
{to reduce the continuum background.}}}

%
%
%

{{Photon} conversions
can
result in large backgrounds
when an electron
is {reconstructed} 
as {a} hadron 
{and still {passes} the electron veto.}
}
{For the $\tau\to ehh'$ modes,
the {$e^-h^+$ and {$h'^-h^+$}} invariant masses 
for {the} $\tau^-\to e^-h^+h'^-$ modes
($e^+h^-$ and $e^+h'^-$ for the $\tau^- \to e^+h^-h'^-$ modes),
{assigning  {the} electron mass to both tracks,}}
{are} required to be greater than 0.2 GeV/$c^2$
to reduce 
photon conversion and other {backgrounds.}

%
%
%

For the $\tau\to \mu hh'$ modes,
a real muon track 
{can come} from a kaon decaying in the CDC ($K^\pm\to\mu^{\pm}\nu_\mu$). 
Therefore, 
we apply a kaon veto, ${\cal{P}}(K/\pi)<0.6$, 
for 
muon candidate tracks if {{the} tag side track} is {a}  hadron
(see Fig.~\ref{fig:pid}~(a)).
{Another {significant} continuum background }
{is} from di-baryon production with a proton on the tag side.
{To suppress this background},
we apply a proton veto, ${\cal{P}}(p/\pi)<0.6$ and ${\cal{P}}(p/K)<0.6$, 
{as shown in  Figs.~\ref{fig:pid} (b) and (c).}

\begin{figure}
\begin{center}
\includegraphics[width=0.9\textwidth]{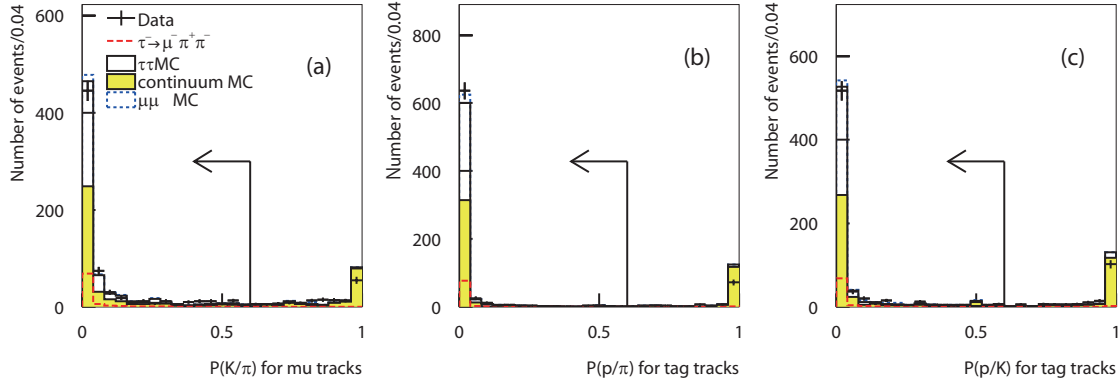}
 \vspace*{-0.5cm}
\caption{
(a) ${\cal{P}}(K/\pi)$ for muon {tracks,} 
(b) ${\cal{P}}(p/\pi)$ and (c) ${\cal{P}}(p/K)$ for  
hadronic tags, 
for $\tau^-\to\mu^-\pi^+\pi^-$ candidate events. 
{{Signal} MC ($\tau^-\to\mu^-\pi^+\pi^-$)
distributions {are} {normalized} arbitrarily 
while the background MC
 distributions are normalized to the data luminosity.}
{{The} selected regions are indicated
 by {arrows}.
}}
\label{fig:pid}
\end{center}
\end{figure}

%
%
%


For the $\tau^-\to\mu^-\pi^+\pi^-$ mode,
{we} reduce the
{$e^+e^-\to\mu^+\mu^-\gamma$ background}
{(with} the
{photon} converting into an electron-positron 
{pair) by requiring}
that the invariant mass of the pion pair, 
evaluated by assuming 
{the} electron mass {for both tracks}, 
{exceed}
0.2 GeV/$c^2$.
{In addition, we require}
the momentum of {a} 
muon in the CM system 
be {less than} 4.0 GeV/$c$ and
$\cos \theta^{\mbox{\rm \tiny CM}}_{\rm tag-miss}<0.97$ 
{if the track is a muon candidate 
{{with} ${\cal{P}}(\mu) >  0.1$.}}}

For the $\tau\to\ell\pi K$ modes,
the background {in} 
{the} signal region 
{is} from $\tau^-\to\pi^-\pi^+\pi^-\nu_{\tau}$ events
{in which both the kaon and lepton 
candidates are misidentified.}
{To reduce {this} background, 
we require the invariant mass of {the}
three charged tracks {on} the signal side, 
$M_{\pi\pi\pi}$, reconstructed by
assigning the pion mass to the tracks, be larger than 1.52 
GeV/$c^2$ (see Fig.\ref{fig:pipipi}).}

%
%

Finally, 
to suppress  backgrounds from generic 
$\tau^+\tau^-$ and continuum {processes,} 
we apply a selection 
based
on the magnitude of 
the missing mass squared $m^2_{\rm{miss}}$.
{The {variable} $m^2_{\rm{miss}}$ is defined as
$E^2_{\rm miss}-p^2_{\rm miss}$,
where $E_{\rm miss} = E_{\rm total}-E_{\rm vis}$,
$E_{\rm total}$ is the sum of the beam energies
and $E_{\rm vis}$ is the total visible energy.}
{{We apply different selection criteria depending on the 
{type of {a} one-prong tag:}
{there should be two emitted neutrinos}
{(a single missing neutrino)}
if the {tagging} 
track is 
{leptonic} 
{(hadronic).}
For the $\tau\to ehh'$, $\mu\pi\pi$ and $\mu KK$ modes,
{we impose the requirements}
$-$1.5 (GeV/$c^2$$)^2$ $<m^2_{\rm{miss}}<1.5$ (GeV/$c^2$$)^2$
for {the} hadronic tag {and}
$-$1.0 (GeV/$c^2$$)^2$ $<m^2_{\rm{miss}}<2.5$ (GeV/$c^2$$)^2$
for {the} leptonic tag.
For the $\tau\to \mu\pi K$ modes,
{where the {residual} background 
{after all selections}
is larger than in other modes,}
we require
the following relation between
$p_{\rm{miss}}$ and $m^2_{\rm{miss}}$:
$p_{\rm{miss}} > -8.0\times m^2_{\rm{miss}}-0.5$
and 
$p_{\rm{miss}} > 8.0\times m^2_{\rm{miss}}-0.5$
for {the} hadronic tag
and 
$p_{\rm{miss}} > -9.0\times m^2_{\rm{miss}}+0.4$
and 
$p_{\rm{miss}} > 1.8\times m^2_{\rm{miss}}-0.4$
for {the} leptonic tag;}
{here} $p_{\rm{miss}}$ is in GeV/$c$ and
$m_{\rm{miss}}$ is in GeV/$c^2$
(see Fig. \ref{fig:miss_mupik}).
Typically, 75\% of the generic $\tau^+\tau^-$ background
{is} removed by 
{these $m^2_{\rm miss}$ requirements}
while 75\% of the signal events  
are retained.

\begin{figure}
\begin{center}
       \resizebox{0.4\textwidth}{0.4\textwidth}{\includegraphics
        {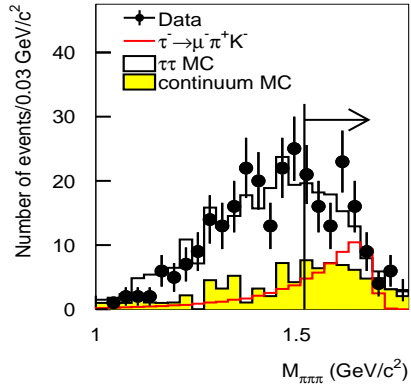}}
\caption{Invariant mass distribution 
of three charged tracks {on} the signal side 
{with}
the pion mass {assigned to each track}
($M_{\pi\pi\pi}$) 
for $\tau^-\to\mu^-\pi^+K^-$ candidate events. 
Signal MC ($\tau^-\to\mu^-\pi^+K^-$)
distributions {are} {normalized} arbitrarily 
while the background MC
distributions are normalized to the data luminosity.
{The} selected regions are indicated
by {the} arrow.
}
\label{fig:pipipi}
\end{center}
\end{figure}

\begin{figure}
\begin{center}
       \resizebox{0.8\textwidth}{0.8\textwidth}{\includegraphics
        {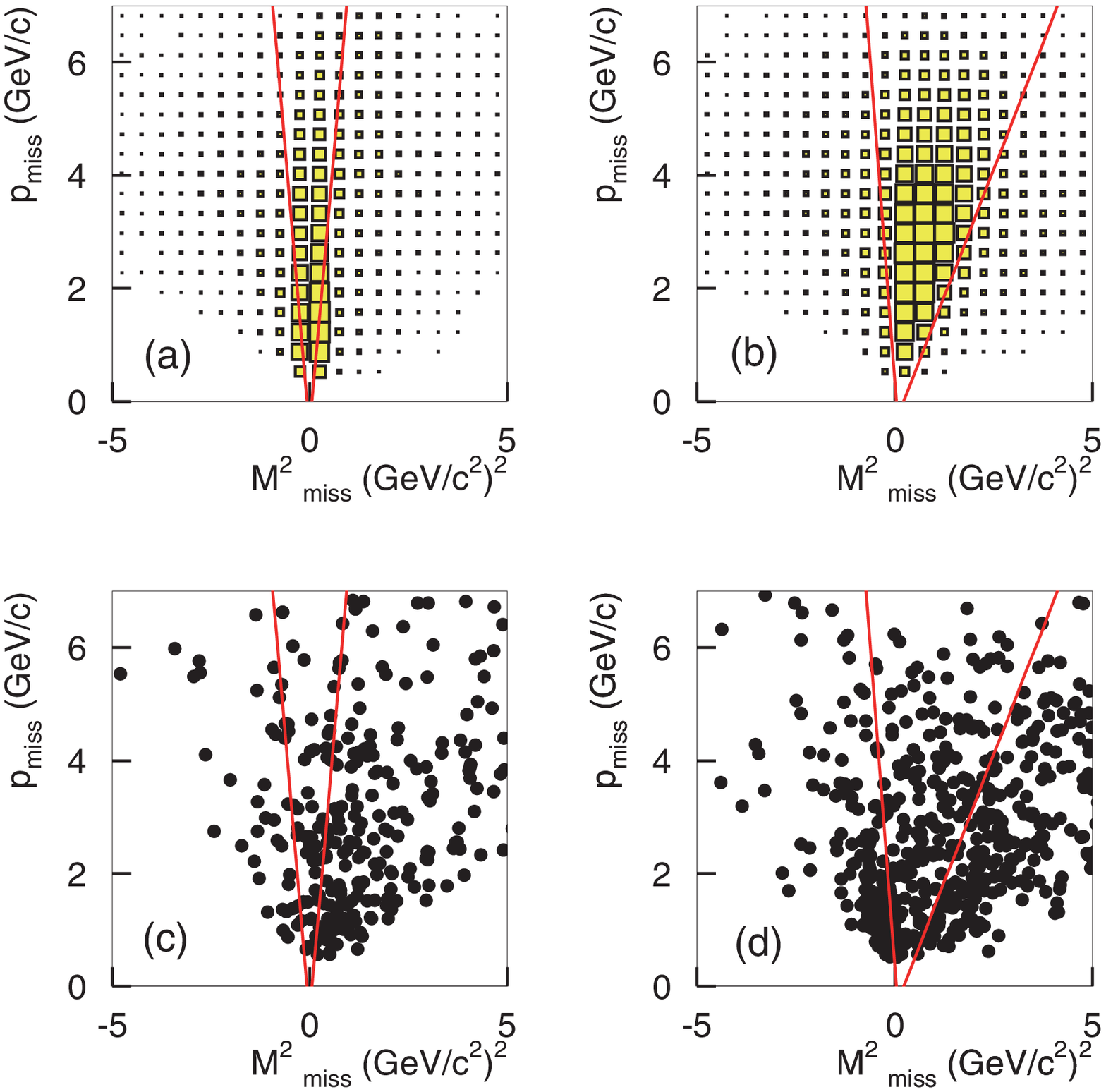}}
\caption{
Scatter plots of
$p_{\rm miss}$
vs.
$m_{\rm miss}^2$ 
for $\tau^-\to\mu^-\pi^+K^-$ modes.
(a) and (c)
show
the signal MC
($\tau^-\to\mu^-\pi^+K^-$)
and generic $\tau^+\tau^-$ MC
distributions,
respectively,
for the hadronic tags
while (b) and (d)  show
the same distributions
for the {leptonic tags.}
Selected regions are indicated by lines.}
\label{fig:miss_mupik}
\end{center}
\end{figure}

\section{Signal and Background Estimation}

For all modes,
the $M_{{\ell hh'}}$ and $\Delta E$  resolutions are 
{obtained}
from fits to the signal MC {distributions,}  
{using}  an asymmetric Gaussian function that takes into account 
{initial-state} radiation.
These {Gaussians} have widths
{as} shown in Table~\ref{tbl:width}.

\begin{table}
\begin{center}
\caption{Summary {of} $M_{\ell hh'}$  
and
$\Delta E$
{resolutions} ($\sigma^{\rm{high/low}}_{M_{\ell hh'}}$ (MeV/$c^2$) 
and 
$\sigma^{\rm{high/low}}_{\Delta E}$ (MeV)).
{Here} $\sigma^{\rm high}$ ($\sigma^{\rm low}$)
{is} the standard deviation
{on the} higher (lower) side of the peak.}
\label{tbl:width}
\vspace*{0.2cm}
\begin{tabular}{c|cccc} \hline\hline
Mode
& $\sigma^{\rm{high}}_{M_{\rm{\ell hh'}}}$
& $\sigma^{\rm{low}}_{M_{\rm{\ell hh'}}}$ 
& $\sigma^{\rm{high}}_{\Delta E}$ 
&  $\sigma^{\rm{low}}_{\Delta E}$ 
 \\ \hline
$\tau^-\to\mu^-\pi^+\pi^-$
& 5.3  &  5.8 & 14.1 & 20.1 \\
$\tau^-\to\mu^+\pi^-\pi^-$
& 5.4   &  5.7 & 14.2 & 20.1 \\

$\tau^-\to e^-\pi^+\pi^-$
& 5.7  &  6.2 & 14.3 & 22.0 \\
$\tau^-\to e^+\pi^-\pi^-$
& 5.6  &  6.3 & 14.4 & 22.3 \\

$\tau^-\to\mu^- K^+K^-$
& 3.4  & 3.6 & 12.9 & 17.2  \\
$\tau^-\to\mu^+ K^-K^-$
& 3.4  &  3.3 & 12.9 & 17.3 \\

$\tau^-\to e^- K^+K^-$
& 4.4  &  4.4 & 13.3 & 19.8 \\
$\tau^-\to e^+ K^-K^-$
& 3.8  &  4.2 & 12.4 & 19.9 \\

$\tau^-\to\mu^-\pi^+K^-$
& 4.4  &  4.8 & 14.2 & 18.8 \\
$\tau^-\to  e^-\pi^+K^-$
& 4.8  &  5.5 & 14.0 & 21.0 \\

$\tau^-\to\mu^-K^+\pi^-$
& 4.6  &  5.1 & 14.3 & 18.7 \\
$\tau^-\to  e^-K^+\pi^-$
& 4.9  &  5.4 & 13.9 & 21.2 \\

$\tau^-\to\mu^+K^-\pi^-$
& 4.5  & 4.7 & 14.7 & 18.6 \\
$\tau^-\to  e^+K^-\pi^-$
& 5.0  &  5.4 & 14.0 & 21.2 \\
\hline\hline
\end{tabular}
\end{center}
\end{table}


To evaluate the branching fractions,
we use  elliptical signal regions
that 
contain 
$\pm3\sigma$
of the MC signal  satisfying all selection criteria.
The ellipse 
is defined as
\begin{equation}
\frac{((M_{\ell hh'}-M_{\ell hh'}^0)\cos\theta-
(\Delta E-\Delta E^0)\sin\theta)^2}
{(3\sigma_{M_{\ell hh'}})^2}
+
\frac{
((M_{\ell hh' }-M_{\ell hh'}^0)\sin\theta+
(\Delta E-\Delta E^0)\cos\theta)^2
}
{(3\sigma_{\rm \Delta E})^2}
=1
\end{equation}
where $M_{\ell hh'}^0$ and $\Delta E^0$ are 
{the coordinates of the} center of the 
ellipse and $\sigma_{M_{\ell hh'}}$ ($\sigma_{\Delta E}$)
is 
{the average of 
$\sigma_{M_{\ell hh'}}^{\rm high}$
and $\sigma_{M_{\ell hh'}}^{\rm low}$
$(\sigma_{{\Delta E}}^{\rm high}$
and $\sigma_{{\Delta E}}^{\rm low})$}
in Table~\ref{tbl:width}.
These elliptical regions are 
determined by 
scanning   
{$M_{\ell hh'}^0$, $\Delta E^0$ and} $\theta$ 
{to maximize the significance {in MC simulation}}
and obtain the highest 
sensitivity.
Table~\ref{tbl:eff2} summarizes the signal
efficiencies for each mode.
We blind the data in the signal region
until all selection criteria are finalized
{to avoid bias.} 

{Figures}~\ref{fig:muhh}  and \ref{fig:ehh} 
show scatter plots 
for data and signal MC samples 
distributed over {a} 
$\pm 20\sigma$ {rectangular region}
in the $M_{\ell hh'}-\Delta E$ plane for the $\tau\to\mu hh'$ 
and $e hh'$ modes, 
respectively.
For the $\tau\to \mu \pi\pi$ {modes,}
{the} dominant background 
{is}
{from}
continuum {processes}
while  
small background {contributions}
{come} from
generic
$\tau^+\tau^-$ events in the $\Delta E<$ 0 GeV and 
$M_{\mu hh'} < $ $m_{\tau}$ region,
which are combinations of a {misreconstructed} 
muon and two {pions.}
For the $\tau\to\mu KK$ {modes,}
{the} dominant background 
{sources are}
continuum 
and
$\tau^+\tau^-$ {processes
with a pion misidentified as a kaon.}
For the $\tau\to \mu \pi K$  modes,
the dominant background {is}
 from
generic
$\tau^+\tau^-$ {decays with}
{combinations of}
{a misidentified muon, a misidentified kaon}
and {a} real pion
from $\tau^-\to \pi^-\pi^+\pi^-\nu$.
If a pion is misidentified as a kaon,
the reconstructed mass from
generic $\tau^+\tau^-$ background 
{can} be greater than the $\tau$ lepton mass 
since the kaon mass is greater than {that} 
of the pion.
For the $\tau\to ehh'$ modes, {the dominant background}
{originates} 
from two-photon processes,
while the {background} {from} 
continuum and generic $\tau^+\tau^-$ {processes} 
is small due to {the} 
low electron fake rate.

{In order to estimate the background in the
signal region, we use the number of
data events observed in the sideband region
inside {a}  $\pm5\sigma_{\Delta E}$ 
{band}
around ${\Delta E} = 0$ GeV
but excluding the signal region.
The events in this band are {projected}  {onto} 
the $M_{\ell hh'}$ axis.
For the $\tau\to\mu\pi\pi$  modes, 
the extrapolation to the signal region 
is performed by 
fitting the sideband of the projected $M_{\mu\pi\pi}$ distribution}
using {the sum of an exponential
and a} first-order polynomial function 
{for} generic $\tau\tau$ and continuum, respectively.
For  the $\tau\to ehh'$,  $\mu KK$ and $\mu\pi K$ 
modes, 
{the background {remaining after all selections}
is small, and}
extrapolation to the signal region 
{assumes that the background is 
linear as a function of $M_{\ell hh'}$.}
The systematic uncertainty due 
to the estimation of the
expected background 
{includes {the}
contributions due to the {statistics} 
of the background sample and the shape of 
the background distribution.}
By varying the assumptions about the background 
shape,
we {verify} that this effect on the systematic uncertainty 
is less than 20\% 
and  is smaller than 
the background statistical error.}
The signal efficiency and 
the number of expected background 
events 
with {its} {uncertainty,} 
obtained by {adding 
statistical and systematic uncertainties in quadrature}
for each {mode,} 
are summarized in Table~\ref{tbl:eff2}.
After estimating the background,  we 
open {the blinded} regions.
We observe one candidate event for 
each of the $\tau^-\to\mu^+\pi^-\pi^-$ and $\mu^-\pi^+K^-$modes,
and no candidate events for {the} other modes.
{The} numbers of events observed in the signal region
are consistent with
{the} expected {background levels}.

The dominant systematic uncertainties
for this analysis
{come
from the resolution {in} $M_{\ell hh'}$ and $\Delta E$ 
and {from} particle identification.}
{We estimate the uncertainties
from resolutions of  $M_{\ell hh'}$ and $\Delta E$
due to the difference between data and MC samples
to be $(3.7-4.8)$\%.}
{The uncertainties due to lepton {identification} 
{are} 
2.2\% and 1.9\% from
{modes with an} electron and {a}
muon, respectively.}
{The uncertainties due to {hadron} {identification} 
{are} 
1.3\% and 1.8\% for pion and kaon {candidates,} 
respectively.}
The uncertainty due to the charged track finding is 
estimated to be {0.35\%} per charged {track;}
therefore, the total uncertainty due to the charged track finding is
{1.4\%.}
The {uncertainty due to integrated} 
luminosity is estimated to be {1.4\%.}
The uncertainties due to the trigger efficiency and MC statistics 
 are negligible 
compared with the other uncertainties.
All these uncertainties are added in quadrature, 
and the total systematic uncertainties for  all modes are
(5.5$-$6.7)\%.

\begin{table}
\begin{center}
\caption{
Summary {of} upper limits for each mode.
{The} table shows
the signal efficiency~($\varepsilon$), 
the number of expected background {events}~($N_{\rm BG}$)
estimated from the  sideband data, 
{the} {total} 
systematic uncertainty~($\sigma_{\rm syst}$),
{the} number of observed events 
in the signal region~($N_{\rm obs}$), 
90\% C.L. upper limit on the number of signal events including 
systematic uncertainties~($s_{90}$) 
and 90\% C.L. upper limit on 
the branching  fraction  (${\cal{B}}$)  
for each individual mode. }
\label{tbl:eff2}
\begin{tabular}{c|cccccc}\hline \hline
Mode &  $\varepsilon$~{(\%)} & 
$N_{\rm BG}$  & $\sigma_{\rm syst}$~{(\%)}
& $N_{\rm obs}$ & $s_{90}$ & 
 ${\cal{B}}~(10^{-8})$ \\ \hline

$\tau^-\to \mu^-\pi^+\pi^- $ & 5.83 & $0.63\pm{0.23}$ & 5.7
 & 0 & 1.87  & 2.1\\
$\tau^-\to \mu^+\pi^-\pi^- $ & 6.55 & $0.33\pm{0.16}$ & 5.6
 & 1 &  4.01  &  3.9\\

$\tau^-\to e^-\pi^+\pi^- $ & 5.45 & $0.55\pm{0.23}$ & 5.7
 & 0 & 1.94  & 2.3\\
$\tau^-\to e^+\pi^-\pi^- $ & 6.56 & $0.37\pm{0.19}$ & 5.5
 & 0 & 2.10  & 2.0\\

$\tau^-\to \mu^-K^+K^- $ & 2.85 & $0.51\pm{0.19}$ & 6.1
 & 0 & 1.97 & 4.4\\
$\tau^-\to \mu^+K^-K^- $ & 2.98 & {$0.25\pm{0.13}$} & 6.2
& 0 & 2.21 & 4.7\\

$\tau^-\to e^-K^+K^- $ & 4.29 & $0.17\pm{0.10}$ & 6.7
 & 0 & 2.29  & 3.4\\
$\tau^-\to e^+K^-K^- $ & 4.64 & $0.06\pm{0.06}$ & 6.5
 & 0 & 2.39  & 3.3\\

$\tau^-\to \mu^-\pi^+K^- $ & 2.72 & $0.72\pm{0.28}$ & 6.2
 & 1 & 3.65  &  8.6\\
$\tau^-\to  e^-\pi^+K^- $ & 3.97 & $0.18\pm{0.13}$ & 6.4
 & 0 & 2.27  & 3.7\\

$\tau^-\to \mu^-K^+\pi^- $ & 2.62 & $0.64\pm{0.23}$ & 5.7
 & 0 & 1.86  & 4.5\\
$\tau^-\to  e^-K^+\pi^- $ & 4.07 & $0.55\pm{0.31}$ & 6.2
 & 0 & 1.97  & 3.1\\

$\tau^-\to \mu^+K^-\pi^- $ & 2.55 & $0.56\pm{0.21}$ & 6.1
 & 0 & 1.93  & 4.8 \\
$\tau^-\to  e^+K^-\pi^- $ & 4.00 & $0.46\pm{0.21}$ & 6.2
 & 0 & 2.03  & 3.2\\








\hline\hline
\end{tabular}
\end{center}
\end{table}

\begin{figure}
\begin{center}
\includegraphics[width=0.9\textwidth]{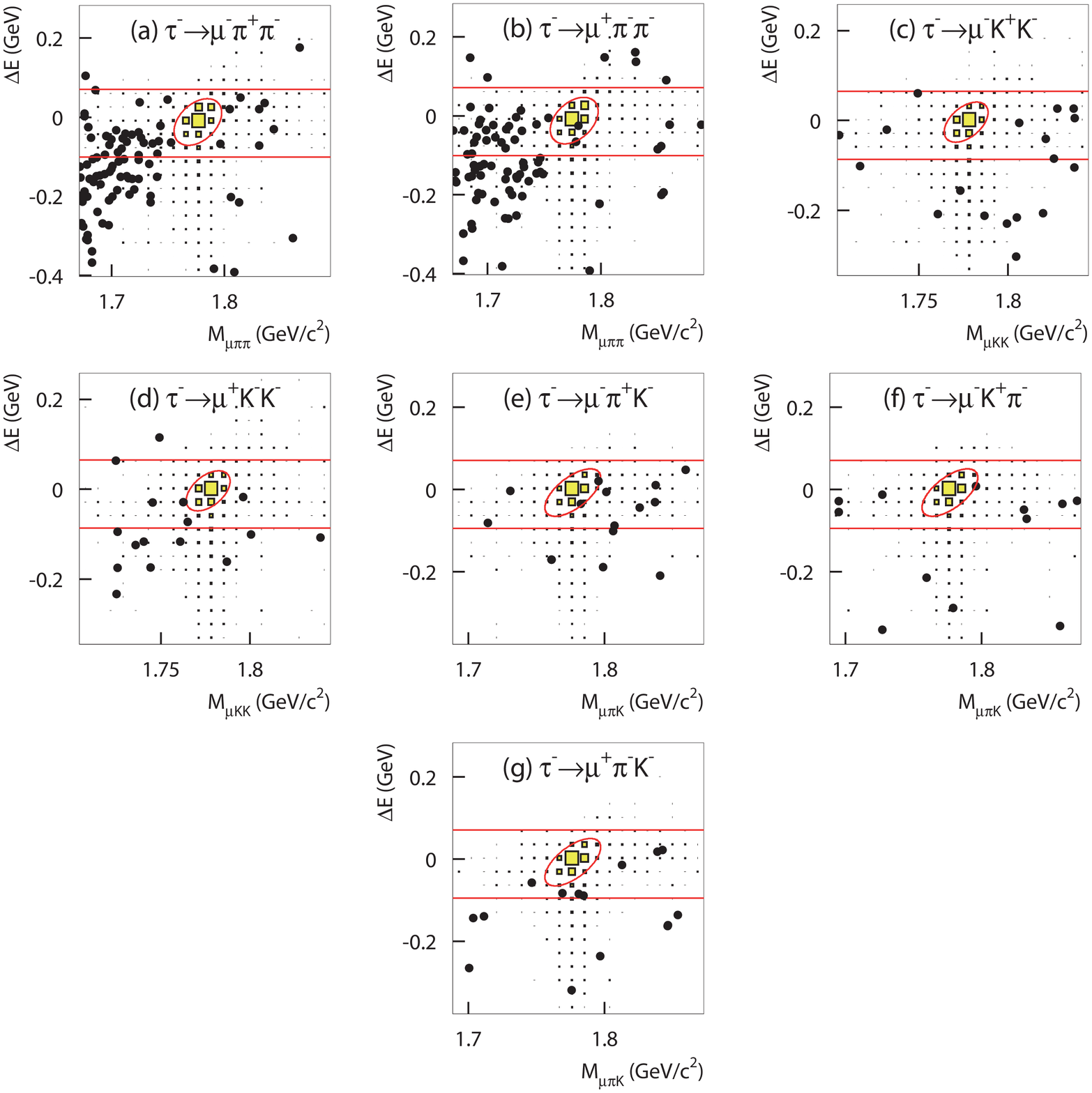}
 \caption{
Scatter plots in the $M_{\ell hh'}$ -- $\Delta{E}$ plane  
{within a}
$\pm 20 \sigma$ {region} for {the}
(a) $\tau^-\rightarrow \mu^-\pi^+\pi^-$,
(b) $\tau^-\rightarrow \mu^+\pi^-\pi^-$,
(c) $\tau^-\rightarrow \mu^-K^+K^-$,
(d) $\tau^-\rightarrow \mu^+K^-K^-$,
(e) $\tau^-\rightarrow \mu^-\pi^+K^-$,
(f) $\tau^-\rightarrow \mu^-K^+\pi^-$,
and
(g) $\tau^-\rightarrow \mu^+\pi^-K^-$
modes.
The data are indicated by the solid circles.
The filled boxes show the MC signal distribution
with arbitrary normalization.
The elliptical signal
{regions}
shown by solid {curves}
are used for evaluating the signal yield.
{The region between {two} horizontal solid lines excluding
the signal region is
used as {a} sideband.}
}
\label{fig:muhh}
\end{center}
\end{figure}

\begin{figure}
\begin{center}
\includegraphics[width=0.9\textwidth]{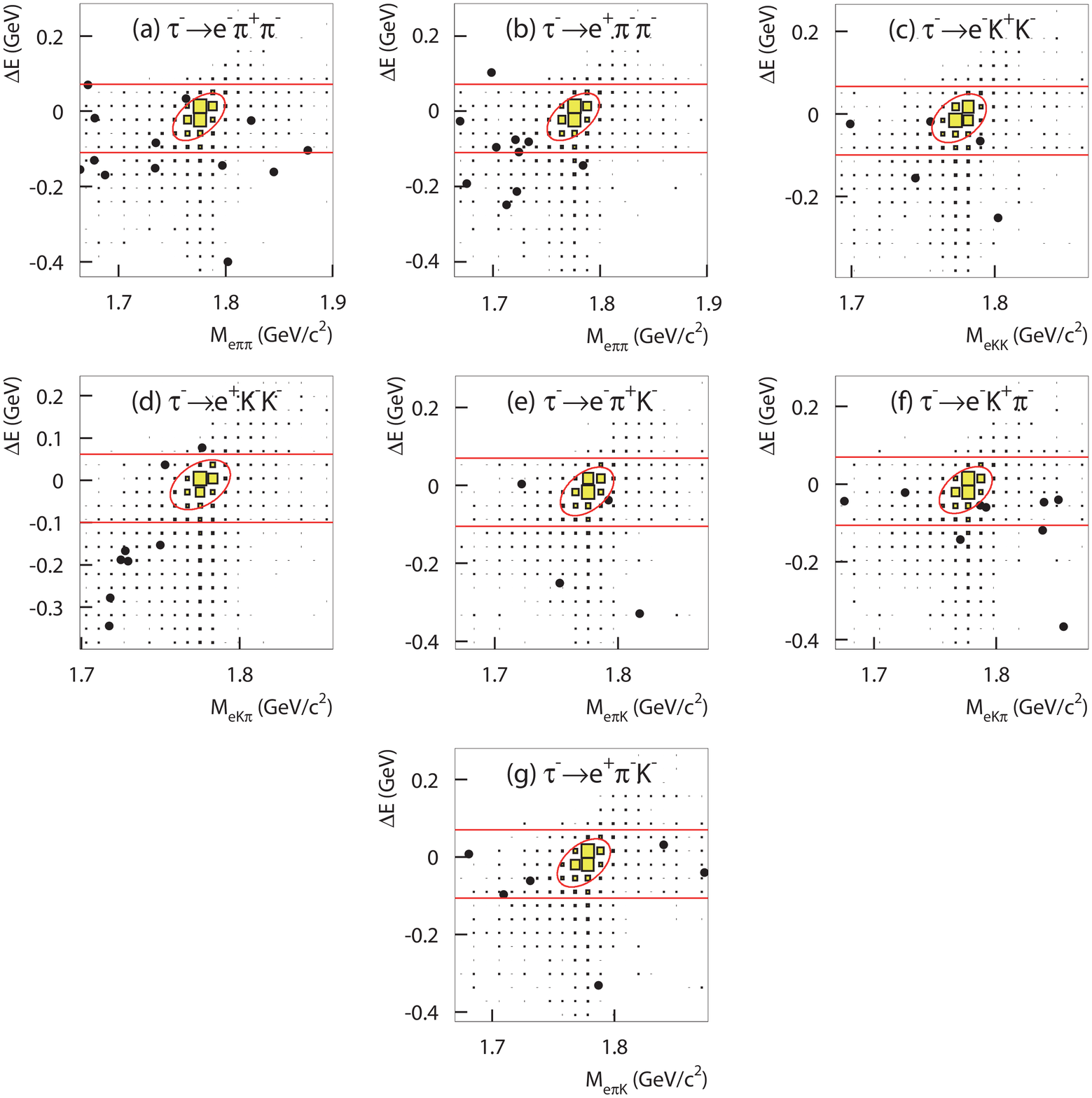}
 \caption{
Scatter plots in the $M_{\ell hh'}$ -- $\Delta{E}$ plane  
{within a}
$\pm 20 \sigma$ {region} for {a}
(a) $\tau^-\rightarrow e^-\pi^+\pi^-$,
(b) $\tau^-\rightarrow e^+\pi^-\pi^-$,
(c) $\tau^-\rightarrow e^-K^+K^-$,
(d) $\tau^-\rightarrow e^+K^-K^-$,
(e) $\tau^-\rightarrow e^-\pi^+K^-$,
(f) $\tau^-\rightarrow e^-K^+\pi^-$,
and
(g) $\tau^-\rightarrow e^+\pi^-K^-$
modes.
The data are indicated by the solid circles.
The filled boxes show the MC signal distribution
with arbitrary normalization.
The elliptical signal
{regions}
shown by  solid {curves}
are used for evaluating the signal yield.
{The region between {two} horizontal solid lines excluding
the signal region is
used as {a} sideband.}
}
\label{fig:ehh}
\end{center}
\end{figure}

\section{Upper Limits on the branching fractions}

Since no statistically significant excess of data over
the expected background in the signal region {is} observed,
we set upper limits on the branching fractions 
of {the} $\tau\to\ell hh'$
{modes using} 
the Feldman-Cousins method~\cite{cite:FC}.
The 90\% C.L. upper limit on the number of signal events 
including  systematic uncertainty~($s_{90}$) is obtained 
{using} the POLE program without conditioning~\cite{pole}
{based on}
the number of expected {background events}, 
{the number of observed events}
and the systematic uncertainty.
The upper limit on the branching fraction ($\cal{B}$) is then given by
\begin{equation}
{{\cal{B}}(\tau\to\ell hh') <
\displaystyle{\frac{s_{90}}{2N_{\tau\tau}\varepsilon{}}}},
\end{equation}
where $N_{\tau\tau}$ is the number of $\tau^+\tau^-$pairs, and 
$\varepsilon$ is the signal efficiency.
{The value {$N_{\tau\tau} =  782\times 10^6$}} is obtained 
from 
{the} integrated luminosity {and}
the cross section of {$\tau$-pair production,} which 
is calculated 
{in 
KKMC~\cite{tautaucs} to be 
$\sigma_{\tau\tau} = 0.919 \pm 0.003$ nb and
$\sigma_{\tau\tau} = 0.875 \pm 0.003$ nb 
for 782 fb$^{-1}$ {at} $\Upsilon(4S)$ 
and 72 fb$^{-1}$ {at} $\Upsilon(5S)$, respectively.}
Table~\ref{tbl:eff2}
{summarizes} 
{the results for}
{all modes.}
{The upper limits for the $\tau\rightarrow ehh'$ modes
are in the range $(2.0-3.7)\times 10^{-8}$
while 
those for the $\tau\rightarrow \mu hh'$ modes
{are in the range}
$(2.1-8.6)\times 10^{-8}$.}
These results improve {upon} 
our previously published upper 
{limits~\cite{lhh_belle} by} factors {of} 
{about} 1.8 on  average.
This improvement results both from 
using {a} 
larger data sample and from
{the} introduction of 
an improved rejection of specific backgrounds, 
such as di-baryon production in the continuum 
for the $\tau\to\mu hh'$ modes,
and {improved}
kinematic event selections, for example, 
{the} $M_{\pi\pi\pi}$ {requirement} for 
{the} $\tau\to \ell \pi K$ modes.

%
%
%
%
%
%

\section{Summary}

We have searched for lepton-flavor
and lepton-number-violating
 $\tau$ decays 
into {a lepton and two charged  mesons}
using 854 fb$^{-1}$ of data.
We obtain  90\% C.L. upper limits 
{on the branching fractions} 
{of $\tau\rightarrow ehh'$ in the range
$(2.0-3.7)\times 10^{-8}$
and 
upper limits on 
$\tau\rightarrow \mu hh'$ in the range
$(2.1-8.6)\times 10^{-8}$.}
These results improve {upon} 
{our} 
previously published upper limits
by factors {of} {about} 1.8 on  average.
These more stringent upper limits can be used
to constrain the space of parameters in various models of new physics.

\section*{Acknowledgments}

%
We thank the KEKB group for the excellent operation of the
accelerator; the KEK cryogenics group for the efficient
operation of the solenoid; and the KEK computer group,
the National Institute of Informatics, and the 
PNNL/EMSL computing group for valuable computing
and SINET4 network support.  We acknowledge support from
the Ministry of Education, Culture, Sports, Science, and
Technology (MEXT) of Japan, the Japan Society for the 
Promotion of Science (JSPS), and the Tau-Lepton Physics 
Research Center of Nagoya University; 
the Australian Research Council and the Australian 
Department of Industry, Innovation, Science and Research;
the National Natural Science Foundation of China under
contract No.~10575109, 10775142, 10875115 and 10825524; 
the Ministry of Education, Youth and Sports of the Czech 
Republic under contract No.~LA10033 and MSM0021620859;
the Department of Science and Technology of India; 
the Istituto Nazionale di Fisica Nucleare of Italy; 
the BK21 and WCU program of the Ministry Education Science and
Technology, National Research Foundation of Korea,
and GSDC of the Korea Institute of Science and Technology Information;
the Polish Ministry of Science and Higher Education;
the Ministry of Education and Science of the Russian
Federation and the Russian Federal Agency for Atomic Energy 
and a Grant of the Russian Foundation for Basic Research 12-02-01032;
the Slovenian Research Agency;  the Swiss
National Science Foundation; the National Science Council
and the Ministry of Education of Taiwan; and the U.S.\
Department of Energy and the National Science Foundation.
This work is supported by a Grant-in-Aid from MEXT for 
Science Research in a Priority Area (``New Development of 
Flavor Physics''), and from JSPS for Creative Scientific 
Research (``Evolution of Tau-lepton Physics'').

\end{document}